\documentclass[letterpaper,twocolumn,showpacs,,superscriptaddress,prb]{revtex4}
\usepackage{graphicx}
\usepackage{amssymb}
\usepackage{amsmath,amsfonts,latexsym}
\usepackage{array,tabularx,color}
\usepackage{dcolumn}               
\usepackage[normalem]{ulem}

\renewcommand{\sout}[1]{}

\newcommand{\vect}[1]{{\mathbf #1}}
\newcommand{\Frac}[2]{\displaystyle\frac{#1}{#2}}

\begin{document}

\title{Coherence properties and luminescence spectra of condensed
  polaritons in CdTe microcavities }

\author{M.~H.~Szyma{\'n}ska}
\affiliation{Clarendon Laboratory, Department of Physics, University of Oxford,
             Parks Road, Oxford, OX1 3PU, UK}
\author{F.~M.~Marchetti}
\affiliation{Rudolf Peierls Centre for Theoretical Physics, 1 Keble
Road, Oxford OX1 3NP, UK}
\author{J.~Keeling}

\affiliation{Cavendish Laboratory, University of Cambridge,
             Madingley Road, Cambridge CB3 0HE, UK}
\author{P.~B.~Littlewood}, 
\affiliation{Cavendish Laboratory, University of Cambridge,
             Madingley Road, Cambridge CB3 0HE, UK}

\date{\today}

\begin{abstract} 
  We analyse the spatial and temporal coherence properties of a
  two-dimensional and finite sized polariton condensate with
  parameters tailored to the recent experiments which have shown
  spontaneous and thermal equilibrium polariton condensation in a CdTe
  microcavity [Kasprzak \emph{et al.}, Nature \textbf{443}, 409
  (2006)].  We obtain a theoretical estimate of the thermal length,
  the lengthscale over which full coherence effectively exists (and
  beyond which power-law decay of correlations in a two-dimensional
  condensate occurs) of the order of $5\mu$m.  In addition, the
  exponential decay of temporal coherence predicted for a finite size
  system is consistent with that found in the experiment.  From our
  analysis of the luminescence spectra of the polariton condensate,
  taking into account pumping and decay, we obtain a dispersionless
  region at small momenta of the order of 4 degrees.
  In addition, we determine the polariton linewidth as a function of
  the pump power.  Finally, we discuss how, by increasing the
  exciton-photon detuning, it is in principle possible to move the
  threshold for condensation from a region of the phase diagram where
  polaritons can be described as a weakly interacting Bose gas to a
  region where instead the composite nature of polaritons becomes
  important.
\end{abstract}


\maketitle

\section{Introduction}
\label{sec:intro}
Since the first observation of polariton bosonic stimulation under non
resonant excitation obtained in 1998 in a CdTe
microcavity~\cite{dang98:prl}, there has been a considerable interest
in realising a polariton Bose-Einstein condensate (see, e.g., the
recent review~\cite{keeling_review07} and references therein).  This
search has been mainly motivated by the expected high transition
temperature of polaritons due to their very light effective
mass~\cite{eastham00:ssc,kavokin03:pla,keeling04:polariton}.  A
spontaneous and thermal equilibrium polariton condensate has been very
recently achieved in a CdTe microcavity~\cite{kasprzak06:nature}.
This was soon followed by similar effects in a GaAs
microcavity~\cite{deng06:eqbm,deng07}.  Other recent progress in
polariton condensation includes studying trapped polaritons in a GaAs
structure under stress~\cite{snoke06:condensation} and room
temperature polariton lasing in GaN~\cite{baumberg06}.

The microcavity polariton system is finite, two-dimensional, decaying,
and interacting, and therefore is expected to differ from the
Bose-Einstein condensation of ideal three-dimensional bosons.  Much
theoretical work involving different modelling has been done to
evaluate the phase diagram for microcavity
polaritons~\cite{eastham00:ssc,kavokin03:pla,malpuech03,keeling04:polariton,marchetti06:prl}
--- for a comprehensive review on the use of and comparison between
different models see Ref.~\cite{keeling_review07}. Already the results
obtained in~\cite{keeling04:polariton,marchetti06:prl}, and adapted to
the CdTe experiment~\cite{kasprzak06:nature}, have shown a very good
agreement between the experimental estimate of the critical density
for condensation obtained from the occupation data --- at a given
temperature and detuning conditions --- and the theoretical estimate
obtained from the lower polariton blue-shift (see the online
supplementary information of Ref.~\cite{kasprzak06:nature}).
Recently, a direct comparison between experimental and theoretical
phase boundaries for condensation of microcavity polaritons in CdTe
under different conditions of cryostat temperature and detuning have
been performed in Ref.~\cite{marchetti07}.  In that work, it was shown
that for a steady state situation, in the presence of pumping and
decay, although the polaritons may display a thermal population, the
presence of pumping and decay may yet have noticeable effects.  In
particular the small discrepancies between the experimental data and
the equilibrium theoretical estimates for a closed system can be
attributed to the effects of pump and decay.

In fact, it has been already
shown~\cite{szymanska06:prl,szymanska06:long} that quantum
condensation is indeed possible in strongly dissipative systems with
continuous pump and decay and that it shares several features
with quantum condensation in a closed system at equilibrium.
In particular, the mechanism of condensation, connected with the
chemical potential reaching the lower polariton (LP) mode, is exactly
the same in closed systems at equilibrium and in open systems with
pump and decay.  In the latter, the role of the chemical
potential is played by the energy at which the non-thermal
distribution diverges. However, it has been
shown~\cite{szymanska06:prl,szymanska06:long} that the presence of
pump and decay can significantly modify the coherence properties of
the condensate. In particular, the power-law decay of temporal and
spatial correlations, caused by the 2D nature of the polariton system,
can be strongly modified by the presence of dissipation.  Finally, it
has been shown~\cite{szymanska06:long} that the finite size of the
system additionally modifies the temporal coherence properties and in
particular it leads to a crossover from a power-law decay of temporal
coherence in an infinite system to an exponential decay in a finite
system.

As well as coherence measurements, changes to the photoluminescence
spectrum can also provide evidence for quantum condensation.  The
photoluminescence spectrum reflects the structure of the normal modes
of the microcavity system, weighted (in thermal equilibrium) by the
bosonic occupation factor.  The structure of collective modes changes
dramatically when microcavity polaritons condense: The lower and upper
polariton modes, which are the eigenmodes of the system in the
non-condensed regime, are now replaced by two new eigenmodes. In
particular, the lower polariton is replaced by the collective
(Goldstone) mode and its quadratic dispersion evolves to a linear
dispersion at small momenta in the closed equilibrium
system~\cite{keeling04:polariton,keeling05,marchetti06:prl}, while it
becomes diffusive at small momenta in the open system with pump and
decay~\cite{szymanska06:prl,wouters05,wouters06,szymanska06:long}.

In this paper we discuss the consequences of these general results for
the properties of coherence and luminescence spectra of microcavity
polaritons in the specific conditions of the CdTe
experiments~\cite{kasprzak06:nature,marchetti07}. In particular, by
considering the closed system, we give estimates of the thermal
length, the lengthscale over which full coherence effectively exists
(and beyond which power-law decay of correlations in a two-dimensional
condensate occurs), and find that its value, of the order of $5\mu$m,
is consistent with the measurements of the spatial decay of the first
order coherence reported in~\cite{kasprzak06:nature}.  We also
consider effects due to pump and decay, and calculate the luminescence
spectra for the conditions of the CdTe experiments and in particular
analyse the size of the diffusive region as well as the size of the
linear regime in the luminescence.  The non-equilibrium formalism,
which accounts for the effects of pump and decay, also allows us to
calculate the homogeneous polariton linewidth as a function of the
pump power.
Finally, in Ref.~\cite{marchetti07} it has been shown that the
experimental data for the phase boundary lie close to the crossover
region between a Berezhinskii-Kosterlitz-Thouless (BKT) transition of
structureless bosons and a regime where instead the boundary is
determined by the long-range nature of the polariton-polariton
interaction. In this paper we show how the increase of the
exciton-photon detuning, which results in a higher transition density
at a given temperature, moves the transition (at a given temperature)
from the BKT weakly interacting Bose boundary to a regime where the
composite nature of polaritons becomes important.

The paper is organised as follows: in Sec~\ref{sec:equ} we analyse
coherence properties on the basis of an equilibrium theory for a
closed system and in Sec.~\ref{sec:nonequ} we discuss how
non-equilibrium and finite size effects modify these results.
Finally, in Sec.~\ref{sec:phase} we analyse the influence of the
exciton-photon detuning on the polariton phase diagram.

\section{Coherence properties at equilibrium}
\label{sec:equ}
The coherence properties of the condensed polariton system are
described by the functional form of the first order coherence as a
function of time and space:
\begin{equation}
  g^{(1)}(t;\vect{r}) = \Frac{\langle \psi^{\dagger} (\vect{r},t)
  \psi^{}(\vect{0},0) \rangle}{\sqrt{\langle
  \psi^{\dagger}(\vect{0},0) \psi^{}(\vect{0},0) \rangle \langle
  \psi^{\dagger}(\vect{r},t) \psi^{}(\vect{r},t) \rangle}}\; ,
\label{eq:gone}
\end{equation}
where $\psi (\vect{r},t)$ is the photon field.  To determine the
  characteristic length- and time-scales in $g^{(1)}(t;\vect{r})$, we
  must therefore evaluate the correlation function in the numerator
  of Eq.~(\ref{eq:gone}), which can be achieved by finding the normal
  modes --- i.e. collective quasi-particle excitations --- of the
  system. We will therefore start our analysis by determining
the spectrum of the collective excitations above the polariton
condensate, making use of a Bose-Fermi model --- tailored
to the parameters characterising the CdTe
microcavities~\cite{kasprzak06:nature,marchetti07} --- which
takes into account the composite nature of polaritons, the quantum well
disorder, and the non-linearities associated with exciton-photon
interactions (for more details on the model see
Refs.~\cite{marchetti06:prl,marchetti06:rrs_long,keeling_review07}).

\subsection{Collective modes of the polariton condensate}
\label{sec:coll}
\begin{figure}
\begin{center}
\includegraphics[width=1\linewidth,angle=0]{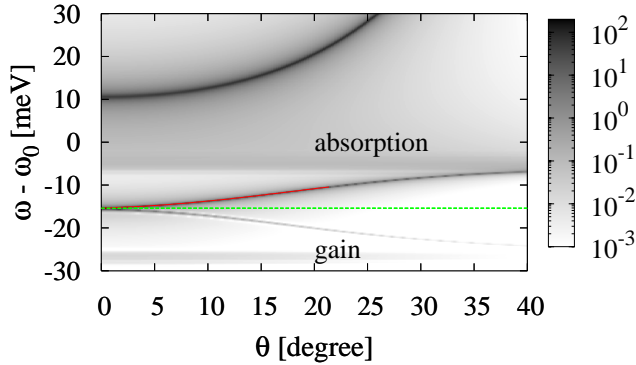}
\end{center}
\caption{\small Contour plot of the spectral weight
  $W(\omega,\vect{p})$ as a function of energy $\omega - \omega_0$,
  where $\omega_0$ indicates the bare cavity photon energy, and
  emission angle $\theta = \sin^{-1} (c|\vect{p}|/\omega_0)$ for a
  closed equilibrium system at temperature $T=19$K, detuning
  $\delta=+6$ meV and a fixed density $n=6.1 \times 10^8$cm$^{-2}$
  (the mean-field critical density for condensation for these values
  of temperature and detuning is given by $n_c=6 \times
  10^7$cm$^{-2}$).  The horizontal dashed (green) line marks the value
  of the chemical potential. The location of the peak for the upper
  branch Goldstone mode is explicitly plotted (red). Its fit to a
  linear dispersion is plotted in Fig.~\ref{fig:linea}.}
\label{fig:golds}
\end{figure}

The spectra of the collective modes of a polariton condensate can be
evaluated by considering the second order fluctuations above the
mean-field approximation (see,
e.g.~\cite{keeling04:polariton,keeling05,marchetti06:prl,marchetti06:rrs_long}).
One can show that, when the system condenses, the lower and upper
polariton modes are not any longer the eigenmodes of the problem and
are instead replaced by new collective modes: The lower polariton
becomes a linear (Goldstone) mode at low momenta, while two new
branches appear below the chemical potential, which are seen as gain
in the spectral weight $W(\omega,\vect{p})$ (see Fig.~\ref{fig:golds}
--- in which the second branch below the chemical potential
emitting at around $\omega-\omega_0 = -41$meV is not shown). Note that
Fig.~\ref{fig:golds} shows the incoherent emission only, without
including the condensate emission.  These features in the spectral
weight will influence the photoluminescence (PL) emission,
\begin{equation}
  P(\omega , \vect{p}) = n_{B} (\omega) W(\omega,\vect{p})\; ,
\end{equation}
which is the product of the spectral weight times the Bose occupation
factor $n_B(\omega)$. The observation of these predicted features in
the photoluminescence would provide very strong evidence for polariton
condensation. However, the emission from below the chemical potential
is suppressed at large angles (see~\cite{marchetti06:rrs_long}), while
the PL emission from modes far above the chemical potential is
suppressed exponentially by the thermal occupation of these modes,
making such features hard to observe. In addition, the strong emission
from the condensate at zero momentum and the frequency corresponding
to the chemical potential, which is not shown in Fig.~\ref{fig:golds},
might mask both the linear behaviour of the collective modes at low
momenta and the emission from below the chemical potential.

Nevertheless, the analysis of the linear behaviour of the Goldstone
mode can give important information on the coherence properties of the
condensate. In particular, the slope of the linear mode is the sound
velocity of the condensate, $c_s$. As it will be shown in the next
section, the sound velocity together with the temperature determines
the lengthscale, the thermal length, for the decay of the spatial
coherence.
  
\begin{figure}
\begin{center}
\includegraphics[width=1\linewidth,angle=0]{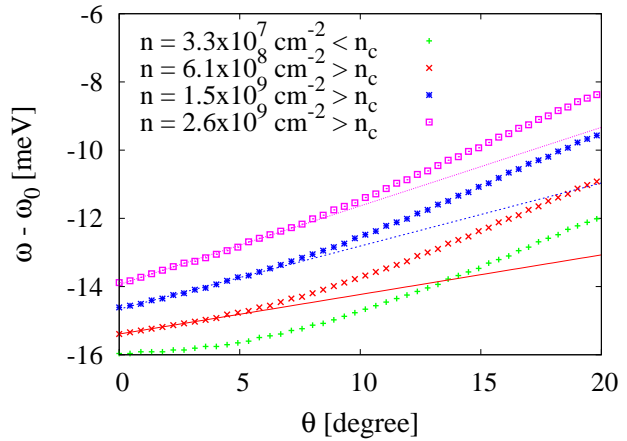}
\end{center}
\caption{\small Dispersion of the lower polariton at a density $n=3
  \times 10^7$cm$^{-2}$ (green) and dispersion of the upper branch of
  the Goldstone mode at $n=6.1 \times 10^8$cm$^{-2}$ (red), $n=1.51
  \times 10^9$cm$^{-2}$ (blue), and $n=2.59 \times 10^9$cm$^{-2}$
  (magenta), for $T=19$K and a detuning $\delta=+6$meV. The fits to a
  linear dispersion of the Goldstone modes are explicitly shown. Note
  that the shift in energy of the emission at zero momentum, which can
  be compared to the experimental one, is for the four curves
  respectively $\delta E = 0.01$meV, $\delta E = 0.55$meV, $\delta E =
  1.30$meV, and $\delta E = 2.04$meV (the shift is measured with
  respect to the energy of the zero momentum lower polariton at
  extremely small densities).}
\label{fig:linea}
\end{figure}

In Fig.~\ref{fig:linea} we plot the upper branch of the Goldstone
mode, obtained from evaluating the locations of the corresponding
peaks as shown in Fig.~\ref{fig:golds}, for increasing values of the
polariton density $n$ ranging from the non-condensed to the condensed
regime. As shown in the picture, the extension of the linear behaviour
grows with increasing density up to approximately $8^{\circ}$ at the
highest density considered. At the same time, the slope of the linear
mode and therefore the sound velocity increases with increasing
density.
%

\subsection{Condensate emission --- decay of spatial and temporal coherence}
\label{sec:conde}
\begin{figure}
\begin{center}
\includegraphics[width=1\linewidth,angle=0]{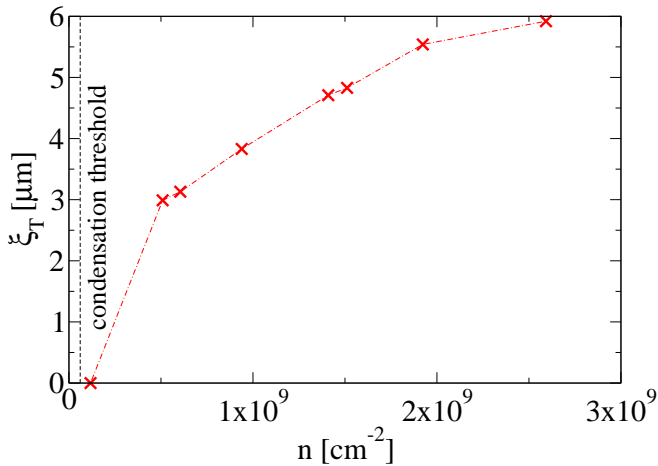}
\end{center}
\caption{\small Dependence of the thermal length $\xi_T$ on the
  polariton density for $T=19$K and a detuning $\delta=+6$meV.}
\label{fig:therm}
\end{figure}

In the condensed state, because there is no restoring force for global
phase fluctuations, the amplitude of phase fluctuations at low momenta
can be large.  However, there is a restoring force for amplitude
fluctuations, and so except near the transition, it is valid to assume
these to be small. Taking the phase fluctuations to all orders and the
amplitude fluctuations to second order, one can calculate the first
order coherence function:
\begin{equation}
  g^{(1)}(t;\vect{r}) = \left[1 +
  \mathcal{O}\left(1/\rho_0\right) \right] \exp\left[ -f(t,\vect{r})
\right]\; .
\label{eq:g}
\end{equation}
In this expression, the terms of order $\mathcal{O}(1/\mathcal{\rho_0})$
arise from including density fluctuations, but are not relevant in
describing the long distance behaviour.  This form asymptotes to the
well-known power law decay of correlations at large times and
distances, for which:
\begin{equation}
  f(t,\vect{r}) = \eta \log \left( \frac{\sqrt{c_s^2 t^2 + r^2}}{\beta
      c_s}\right) \; .
\label{eq:9}
\end{equation} 
Here, $c_s$ is the velocity of the Goldstone mode, $\rho_0$ is the
mean-field estimate of the condensate density and $\eta = m k_B T /
(2\pi \rho_0)$.  The logarithmic dependence on the spatial coordinate
$r$ reflects the expected BKT decay of correlations for a
two-dimensional system.  Note that since the asymptotic form of the
spatial decay is power-law, and not exponential, there is strictly no
characteristic lengthscale of the decay, so the coherence length is
not a well defined quantity.  However, there does exist a lengthscale
--- the thermal length --- which describes the distances at which the
long range asymptotic behaviour begins to be relevant; below this
lengthscale, correlations are short range, and approximately
independent of distance.  This thermal length is given by:
\begin{equation}
  \xi_{T} = 2\pi \Frac{c_s}{k_B T} \; .
\label{eq:thlen}
\end{equation}
Substituting the values of the velocity of the Goldstone mode $c_s$
determined in Section~\ref{sec:coll} we obtain thermal lengths given
by $\xi_T = 3.13\mu$m, $4.83\mu$m, and $5.92\mu$m respectively for
polariton densities of $n=6.1 \times 10^8$cm$^{-2}$, $1.51 \times
10^9$cm$^{-2}$, and $2.59 \times 10^9$cm$^{-2}$ for $T=19$K and a
detuning $\delta=+6$meV.  The dependence of the thermal length on the
density is shown in Fig.~\ref{fig:therm}.
Experiments on coherence in CdTe reported a contrast of interference
fringes up to 5\% below the transition and up to 45\% above the
transition~\cite{kasprzak06:nature}. The 45\% contrast has been
measured up to a distance of roughly 6 $\mu$m after which the contrast
drops to 30\% up to a distance of roughly $12\mu$m.  Our estimates of
the thermal length, of the order of $5\mu$m, are consistent with these
measurements.  Unfortunately due to a very large spatial inhomogeneity
of the condensate caused by large photonic disorder it was not
possible in that experiment to examine a functional dependence of the
decay of the correlation as a function of the distance. Spatial
coherence has also been recently measured in GaAs
systems~\cite{deng07}; such systems appear to show less spatial
inhomogeneity, and so might allow fuller investigation of the decay of
correlations at long distances.  However, the data presented in
Ref.~\cite{deng07} shows measurements of $g^{(1)}(0;\vect{r})$ at only
six different separations, ranging between $1.3\mu$m and $8\mu$m.
According to the analysis in that paper, the data are adequately
described over that range by modelling the system as a degenerate Bose
gas, without considering any changes to the dispersion of the bosons.
Such a model is in effect a model of non-interacting Bosons, and so
does not describe power law decay of correlations.

\section{Influence of pump, decay and finite size on coherence properties}
\label{sec:nonequ}
Since the effects of pump and decay in the polariton system are large
compared to other relevant energy scales, such as temperature (one
can, e.g., compare the homogeneous linewidth of polaritons, of roughly
$1$meV with the characteristic temperature $\sim 20$K$\sim 1.7$meV),
we expect pump and decay to modify the coherence properties of the
polariton condensate. We have addressed these issues by looking at the
spontaneous condensation for a system, coupled to external baths,
representing the pumping and decay
mechanisms~\cite{szymanska06:prl,szymanska06:long}. We have shown
that, even when the polariton system is characterised by a thermal
distribution, the presence of pumping and decay significantly modify
the spectra of collective excitations.  In particular, the low energy
phase modes become diffusive at small
momenta~\cite{szymanska06:prl,wouters05,wouters06,szymanska06:long},
leading to correlation functions --- and thus condensate line-shape
--- that differ both from an isolated equilibrium BEC and from those
for phase diffusion of a single laser mode. Here we give estimates of
the size of the diffusive region for conditions close to those of
the CdTe experiments.

\subsection{Collective modes of the polariton condensate in presence
  of pump and decay}
\begin{figure}
\begin{center}
\includegraphics[width=1\linewidth,angle=0]{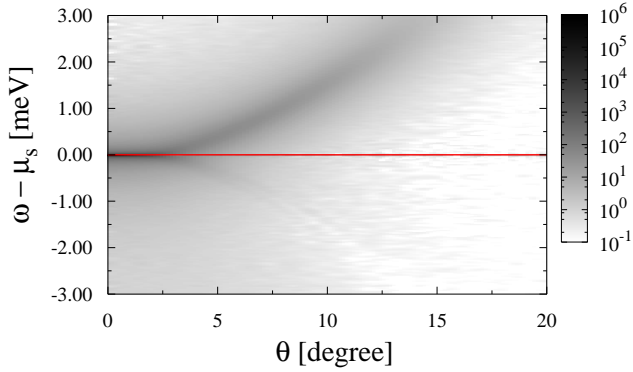}
\end{center}
\caption{\small Condensed luminescence spectra for the photon decay
  rate of $1$ps ($\sim 0.49$meV) and the parameter $\gamma$,
  representing the coupling strength to the pumping baths,
  $\gamma$=1.3meV.  The frequency is shown with respect to the
  condensate frequency. The flat region is around 3 degrees. }
\label{fig:non-eq-lumsmall}
\end{figure}

In the normal state the fluctuation spectrum shows the usual polariton
branches, which are now also homogeneously broadened due to pumping
and decay. In the condensed state, however, the collective modes
have the following energy $\omega$ vs. momentum $p$ form:
\begin{equation}
  \omega = - i x \pm i \sqrt{x^2 - c_s^2 p^2} \; ,
\end{equation}
and are thus diffusive, rather than dispersive for $p\le x/c_s$.
Here, the parameter $x$ is a non-linear function of the pumping and
decay strength and determines the linewidth of the Goldstone mode.
Similarly to an equilibrium picture, the structure of the collective
modes will be reflected in the PL which is a product of the spectral
weight and the occupation function. In Figure
\ref{fig:non-eq-lumsmall} we plot the PL spectra for the parameters
characterising the recent experiments on CdTe
\cite{kasprzak06:nature,marchetti07}.
It can be seen that in its low frequency and momentum part, the main
feature of the spectrum is a flat region followed by a dispersive mode
which then approaches the LP spectrum.  As in the closed system, the
coupling of particle- and hole-like bosonic excitations means that in
the condensed state, the same mode structure can also be seen below
the chemical potential.  The spectral weight of this mode is however
much weaker and so it is less visible in the luminescence in
Fig.~\ref{fig:non-eq-lumsmall}.  Our numerical analysis shows that the
range of the flat region depends mainly on the photon decay rate; but
also depends weakly on the pump parameter $\gamma$ such that the flat
region should increase slightly with increasing pump power which is
consistent with the experiment~\cite{kasprzak06:thesis}. For the
photon decay rate of $1$ps ($\sim 0.49$meV) the size of the flat
region extends up to around 5 degrees (depending on the coupling
strength $\gamma$). There is also an alternative explanation that the
observed flattening above the condensation transition is mainly due to
the finite size effects~\cite{kasprzak07:pol}.  Note that since the
size of the flat region here is of similar order to the range over
which the Goldstone mode is linear (see
Figure~\ref{fig:non-eq-lumsmall}), it is therefore possible that, at
least close to the transition, the dispersion will cross directly from
diffusive to quadratic, without a linear part.

\subsection{Decay of temporal coherence}
Similarly to an equilibrium picture, in order to examine the coherence
properties of a condensed system, one needs to determine the
field-field correlation functions including the phase fluctuations to
all orders and to determine the analogue of expression~\eqref{eq:g}
for a dissipative system. This approach has several advantages.
Firstly it naturally includes both condensate and non-condensate
luminescence in the same formalism, and (in the finite system)
provides a linewidth for the condensed part.  Secondly it recovers the
correct power-law behaviour of occupation of modes in momentum space
when integrated over frequencies. It has been shown
\cite{szymanska06:prl,szymanska06:long} that for the infinite system
with pump and decay the power-law which governs the decay of
correlations in a 2D system changes to:
\begin{equation}
  f(t,\vect{r}) \simeq
  \begin{cases}
    \Frac{\eta'}{2} \log \Frac{c_s^2 t}{x \xi_c^2}
    & \text{if $r\simeq 0$, $t \to \infty$,} \\[1.5em]
    \eta' \log \Frac{r}{\xi_c} & \text{if $r \to \infty$, $t \simeq
      0$.}
  \end{cases}\; ,
\label{eq:limiting-decays}  
\end{equation}
where $\eta^{\prime}$, in distinction to $\eta$ in
  Eq.~(\ref{eq:9}), depends on $x$ as well as temperature and
  condensate density, and $\xi_c$ is a characteristic length scale
for the non-equilibrium occupation function of polaritons,
given by $\xi_c\propto c_s/E$, where $E$ is a characteristic
energy scale of the polaritons' distribution.  Thus, there
is still power-law decay, but due to pumping and decay the powers for
temporal and spatial decay do not match. In the case of
systems with strong pumping and decay, but where the
distribution function is close to thermal, as in the recent
experiments on CdTe~\cite{kasprzak06:nature}, then $E\simeq k_B
  T$ and so
\begin{equation}
  \label{eq:noneqb-cutoff}
  \xi_c= 2\pi \Frac{c_s}{k_BT} \equiv \xi_T \; .
\end{equation}
Therefore, if the system, despite the presence of pump and decay, is
able to thermalize, the thermal length is not strongly affected by the
presence of pump and decay and its expression coincides with
that of a closed system~\eqref{eq:thlen}.

\begin{figure}
\begin{center}
\includegraphics[width=1\linewidth,angle=0]{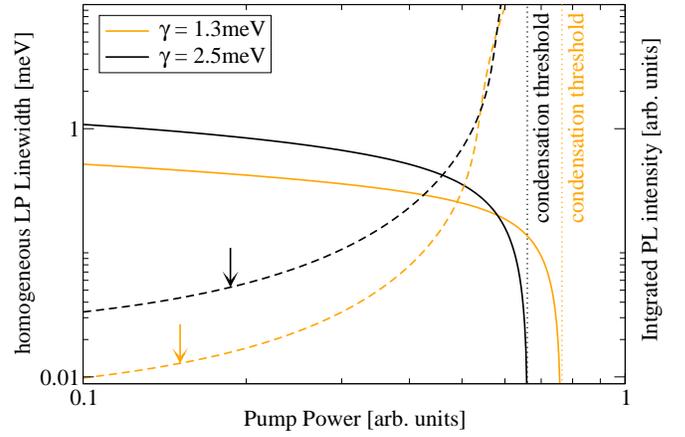}
\end{center}
\caption{\small Calculated homogeneous linewidth of the zero momentum
  lower polariton (solid line) and the integrated zero momentum PL
  intensity as a function of the pump intensity for two different
  dephasing parameters $\gamma$. The decay rate of the photon is
  determined from the homogeneous photon linewidth, measured to be
  around 1meV. The threshold for non-linear emission is explicitly
  shown.}
\label{fig:linew}
\end{figure}

In order to understand the temporal coherence
measurements~\cite{kasprzak06:nature} which show exponential decay,
it turns out to be necessary to address the influence of
finite size on the expression~\eqref{eq:g}. A detailed analysis of
this can be found in reference~\cite{szymanska06:long}. Summarising,
in the finite condensed system the energy level spacing is given by
$\Delta_\phi = c_s/R$ [note this differs from the single
particle level spacing relevant for the uncondensed regime
$\Delta_{\rm{s.p.}}  = 1/ (2mR^2)$]. With this level spacing it has
been shown~\cite{szymanska06:long} that the function which controls
the decay of coherence can be approximately written as
\begin{equation}
  f(t,\vect{r}=0) \propto \frac{1}{x}\left[\frac{t}{2x}+
    \frac{1}{\Delta_{\phi}^2}
    \log\left((k_B T)^2 \Frac{t}{x}\right)\right]\; . 
\end{equation}
In this expression, we have used the assumption of a thermal
  distribution to write $k_BT$ for the high energy cutoff
  $c_s/\xi_{c}$ as in Eq.~(\ref{eq:noneqb-cutoff}).  The first term,
which dominates at large times, gives exponential decay of
correlations whereas the second term gives the power-law decay
characteristic for infinite 2D systems. The relative importance of
these two terms depends on the system size and temperature (i.e how
deep the system is into the condensed regime). If the system is large
or is close to the phase boundary i.e where $k_{B} T \gg
  \Delta_\phi = c_s/R$ the second term (i.e power-law decay)
dominates at short times and it crosses to  exponential decay only
at later times.  Rearranging this condition it translates to $R \gg
2\pi c_s/k_B T = \xi_T$ which says that the system size is much larger
than the thermal length. However if the system is small or deep inside
the condensed region and so $k_BT \ll \Delta_\phi = c_s/R$ the phase
fluctuations associated with 2D nature of the condensate are frozen
out and the first term dominates giving an exponential decay at
  all times.  This conditions translates to $R \ll 2\pi c_s/k_B T =
\xi_T$. Thus spatial coherence over the whole
system size implies the exponential decay of temporal coherence
(however not vice-versa).

\begin{figure}
\begin{center}
\includegraphics[width=1\linewidth,angle=0]{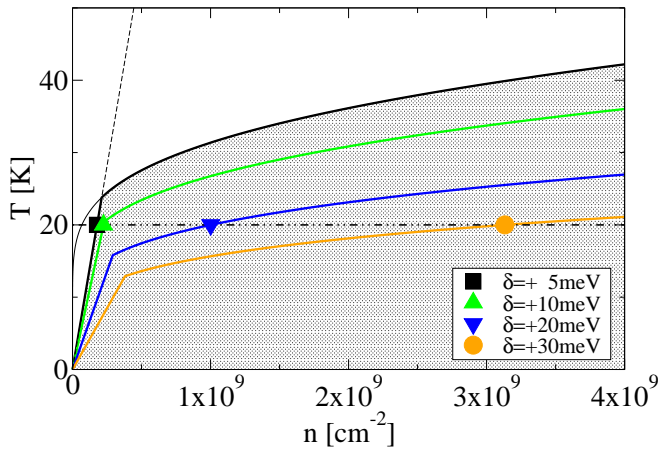}
\end{center}
\caption{\small Dependence of the polariton phase diagram on the
  detuning $\delta$ of the exciton mode below the bottom of the
    photon band. By fixing the polariton effective temperature at
  around $20$K, the transition to the condensed phase (symbols) moves
  from the linear regime of the phase diagram to the regime where the
  boundary is determined by the long-range nature of the
  polariton-polariton interaction.}
\label{fig:phased}
\end{figure}
%
\subsection{Linewidth in the normal state}
\label{sec:linew}
In order to determine the homogeneous contribution to the polariton
linewidth, i.e. that part due to pumping and decay as opposed to the
linewidth resulting from structural disorder, we must explicitly
consider an open system with incoherent pumping and decay of photons.
Here, we focus only on the homogeneous linewidth of the normal state
on approaching the transition.  Following
Refs.~\cite{szymanska06:prl,szymanska06:long}, the pumping process is
described by two parameters: The occupation of the pumping bath which
describes the pumping intensity, and the coupling constant $\gamma$
between the system and the pumping bath, which gives rise to
dephasing.  Figure~\ref{fig:linew} shows the LP linewidth (solid
lines) and the energy integrated zero momentum PL intensity (dashed
lines) as a function of the pumping strength for two different values
of $\gamma$. In both cases, the curves are plotted up to the pump
power value corresponding to condensation.

It can clearly be seen that, at pumping powers below the phase
transition, the intensity of photoluminescence already starts to
increase non-linearly. Such non-linear effects in advance of the
condensation transition are not surprising, since a second-order phase
transition is associated with divergent susceptibilities. In addition,
the homogeneous linewidth decreases as one approaches the transition,
as the gain due to pumping balances the loss due to decay.  Therefore,
the pump power at which a minimum of linewidth is observed, which
coincides with the appearance of temporal coherence, may provide a
better indication of the phase boundary than is provided by the
maximum nonlinearity of the photoluminescence, i.e. the threshold
condition.

\section{Exciton photon detuning}
\label{sec:phase}
In this last section we discuss some aspects of the polariton phase
diagram and the possibility of experimentally exploring different
parts of it. As already mentioned in the introduction, there has been
a recent investigation~\cite{marchetti07} of the direct comparison
between experimental and theoretical phase boundaries for condensation
of polaritons in a CdTe microcavity. This work has shown that the
current experimental data for the phase boundary lie close to the
crossover between a BKT transition of structureless bosons (low
density part of the diagram shown in Fig.~\ref{fig:phased}) and a
regime where instead the phase boundary is characterised by the
long-range nature of the polariton-polariton interaction and where
therefore the composite nature of polaritons matters (a region where
the dependence of the critical temperature on the density is slower
then linear).  Those results therefore suggest that polariton
condensation departs from the weakly interacting boson picture.

In that experiment~\cite{marchetti07}, however, it has proven
particularly difficult to change substantially the effective
temperature of polaritons by changing that of the lattice (i.e. the
cryostat temperature): Because of the short polariton lifetime, the
polariton temperature is decoupled from that of the lattice.  It has
therefore not been possible to explore parts of the phase diagram
other than the crossover region by means of changing temperature.
However, if the effective temperature does not change much, then one
can shift the boundary from the structureless boson to the long-range
interaction part of the phase diagram, by changing the detuning
between the photon and the exciton (see Fig.~\ref{fig:phased}).

The change to the phase boundary due to detuning can be simply
explained in the low density (weakly interacting Bose gas) limit, as
the increase of the polariton effective mass with the detuning --- in
this regime one can show that $k_BT_c \propto n/m_{\text{pol}}$
and neglecting exciton dispersion one may write, $m_{\text{pol}} =
2m_{\text{photon}} /[1 - \delta/\sqrt{\delta^2 + \Omega_R^2}]$, in
which $m_{\text{photon}}$ is the photon mass, and $\Omega_R$ the
Rabi-splitting at zero detuning, and zero-density.  In the
high density, long-range interaction part of the phase diagram
the decrease of the critical temperature with the detuning at a fixed
density is due to two mechanisms: a loss of coherence in the
system as, when increasing the detuning, the lower polariton becomes
less photon-like; and the decrease of the effective exciton-photon
coupling, which controls the critical temperature at higher
densities. In Fig.~\ref{fig:phased} we show that in order to see a
significant move of the condensate threshold from the
low-density to the high-density part of the phase
diagram, one has to go at least to positive detunings larger that
$20$meV.  Unfortunately, this has proven to be a challenge
experimentally, and current experiments in CdTe allow one to reach a
maximum detuning of roughly $12$meV~\cite{marchetti07}.

\section{Conclusions}
\label{sec:concl}
To summarise, we have analysed the spatial and temporal coherence
properties of a two-dimensional, finite, and decaying condensate with
parameters tailored to the recent experiments on CdTe
microcavities. We have shown that the theoretical estimate of the
thermal length (over which there is no decay of coherence) of up to 6
$\mu$m, and the exponential decay of temporal coherence are consistent
with those found in experiment. We have also estimated the size of the
dispersionless (flat) region in the PL --- a manifestation of the
diffusive nature of the Goldstone mode --- to be around 5 degrees.
This result suggests that the flattening of the polariton dispersion
above the transition may be attributed to the dispersionless nature of
the Goldstone mode. Since at current experimental conditions the
linear part of the dispersion is of a similar size to the diffusive
part it is likely that the flat region will cross directly to the
quadratic dispersion and that the linear part will not be visible.  In
order to see linear dispersion, it would be necessary to decrease the
size of the diffusive regime, which would require an improvement in
the quality of the cavity mirrors.  Finally we analyse the dependence
of the phase diagram on exciton-photon detuning and suggest that going
to higher positive detunings might provide a means of exploring
different parts of the phase diagram.

\paragraph*{Acknowledgements}
We are grateful to L. S. Dang and J. Kasprzak for stimulating
discussions.  F.M.M. and M.H.S. would like to acknowledge financial
support from EPSRC. J.K. would like to acknowledge financial support
from Pembroke College Cambridge.

\bibliography{ssc}

\end{document}